# A Two-Dimensional Electron Gas Based on a 5*s* Oxide with High Room-Temperature Mobility and Strain Sensitivity

Zexin Feng[a], Peixin Qin[a], Yali Yang[b], Han Yan[a], Huixin Guo[a], Xiaoning Wang[a], Xiaorong Zhou[a], Yuyan Han[c], Jiabao Yi[d], Dongchen Qi[e,f], Xiaojiang Yu[g], Mark B. H. Breese[g,h], Xin Zhang[a], Haojiang Wu[a], Hongyu Chen[a], Hongjun Xiang[b,i], Chengbao Jiang[a], Zhiqi Liu[a,*]

[a] *School of Materials Science and Engineering, Beihang University, Beijing 100191, China*

[b] *Key Laboratory of Computational Physical Sciences (Ministry of Education), State Key Laboratory of Surface Physics, and Department of Physics, Fudan University, Shanghai 200433, China*

[c] *The Anhui Key Laboratory of Condensed Matter Physics at Extreme Conditions, High Magnetic Field Laboratory, Chinese Academy of Sciences (CAS), Hefei, Anhui Province 230031, China*

[d] *Global Innovative Centre for Advanced Nanomaterials School of Engineering, The University of Newcastle, Callaghan, NSW 2308, Australia*

[e] *School of Chemistry, Physics and Mechanical Engineering, Queensland University of Technology, Brisbane, Queensland 4000, Australia*

[f] *Department of Chemistry and Physics, La Trobe Institute for Molecular Science, La Trobe University, Melbourne, Victoria 3086, Australia*

[g] *Singapore Synchrotron Light Source, National University of Singapore, 5 Research Link, Singapore 117603, Singapore*

[h] *Department of Physics, National University of Singapore, Singapore 117542, Singapore*

[i] *Collaborative Innovation Center of Advanced Microstructures, Nanjing 210093, China*

*email: zhiqi@buaa.edu.cn

**ABSTRACT**

The coupling of optical and electronic degrees of freedom together with quantum confinement in low-dimensional electron systems is particularly interesting for achieving exotic functionalities in strongly correlated oxide electronics. Recently, high room-temperature mobility has been achieved for a large bandgap transparent oxide – $BaSnO_3$ upon extrinsic La or Sb doping, which has excited significant research attention. In this work, we report the realization of a two-dimensional electron gas (2DEG) on the surface of transparent $BaSnO_3$ via oxygen vacancy creation, which exhibits a high carrier density of ~$7.72 \times 10^{14}$ cm$^{-2}$ and a high room-temperature mobility of ~18 cm$^2 \cdot$V$^{-1} \cdot$s$^{-1}$. Such a 2DEG is rather sensitive to strain and a less than 0.1% in-plane biaxial compressive strain leads to a giant resistance enhancement of ~350% (more than 540 k$\Omega$/□) at room temperature. Thus, this work creates a new path to exploring the physics of low-dimensional oxide electronics and devices applicable at room temperature.

**KEYWORDS:** two-dimensional electron gas; 5$s$ oxide; $BaSnO_3$; piezoelectric strain modulation; memory devices

1. Introduction

As a result of significantly reduced phonon scattering and the substantial dielectric screening of electric potentials of ionized defects resulting from the giant dielectric constant of $SrTiO_3$ (STO) [1], the mobility of electrons in STO Ti 3$d$ channels, such as the 2DEG in STO-based interface systems [2,3], rises rapidly with lowering temperature following power laws and thus is ultrahigh at low temperatures. However, in addition to the enhanced thermal fluctuation, due to the itinerant nature of Ti 3*d* orbits and the interband scattering originating from the band degeneracy lift by the oxygen octahedron crystal field, the mobility of Ti 3*d* electrons is relatively low at room temperature, typically on the order of ~1 cm$^2$·V$^{-1}$·s$^{-1}$ [2,4,5]. This largely prevents STO-based 2DEG systems from practical oxide electronic device applications.

Recently, BaSnO$_3$ (BSO), a cubic wide bandgap perovskite oxide with a lattice constant of *a* = 4.116 Å and an indirect bandgap of ~3.1 eV [6], has received great attention [7] in the field of oxide electronics owing to its highly mobile conducting channel at room temperature after La [8] or Sb doping [9]. For example, the first study on highly mobile BSO by Kim *et al.* [8] reported a room-temperature mobility of 320 cm$^2$·V$^{-1}$·s$^{-1}$ for La-doped BSO single crystals and later they reported a room-temperature mobility of 79.4 cm$^2$·V$^{-1}$·s$^{-1}$ for Sb-doped BSO single crystals [9]. Meanwhile, high room-temperature mobility was achieved in La-doped BSO thin films as well. Specifically, Kim *et al.* [8] using pulsed laser deposition, Prakash *et al.* [10] by hybrid molecular beam epitaxy, Raghavan *et al.* [11] via a modified oxide molecular beam epitaxy approach and Paik *et al.* [12] with molecular beam epitaxy obtained 70, 120, 150 and 183 cm$^2$·V$^{-1}$·s$^{-1}$, respectively.

In BSO, the electronic configuration of the pivotal element Sn $4d^{10}5s^25p^2$ becomes $4d^{10}5s^05p^0$ after oxidation. As the 4*d* orbit is full, it does not contribute to the conduction band of BSO, which instead is mainly derived from the Sn 5*s* orbit [13]. The densities of states of the Sn 5*p* orbit and O *s*,*p* orbits that contribute to the conduction band of BSO exist at least 1.5 eV away from the bottom of the conduction band [13] and thus these orbits are not active at room temperature. In this highly mobile 5*s* oxide, if a two-dimensional electron channel could be created similar to the 2DEG at the LaAlO$_3$/STO (LAO/STO) interface [2], it would be highly interesting for low-dimensional physics study and electronic device applications.

There are at least two possible means to realizing a 2DEG on the surface of an insulating oxide, polar discontinuity and oxygen vacancies. The first mechanism needs strict band alignment conditions [14] in addition to different nature of polarity [15] to enable the electronic reconstruction on the surface of an insulating oxide. In contrast, the second mechanism could be easily achieved by redox reactions [5,16] once the oxygen vacancy donor level is shallow enough in the insulating oxide. For example, the deposition of an addition LAO layer at high [5,17] or even room temperature [5,16] in a reduced oxygen atmosphere can effectively create oxygen vacancies on the oxide surfaces due to the high chemical affinity of Al atoms to oxygen.

Although early theoretical calculations implied that native point defects such as oxygen vacancies cannot serve as shallow donors for BSO to make it conducting [18], recent growth of BSO thin films by sputtering has suggested that oxygen vacancies could realize *n*-type doping in BSO [19,20]. Hence, it is possible to realize highly mobile 2DEG on the surface of BSO via the oxygen vacancy creation.

## 2. Experimental methods

The 320-nm-thick $BaSnO_3$(BSO) thin films were fabricated onto (001)-oriented double-side-polished MgO single-crystal substrates at different oxygen partial pressures ranging from $10^{-1}$ to $10^{-7}$ Torr under 800 °C by pulsed laser deposition with a based pressure of $1.5 \times 10^{-8}$ Torr (Shenyang Baijujie Scientific Instrument Co., Ltd). The laser fluence and the repetition rate was ~1.6 J/cm$^2$ and 10 Hz, respectively. The highly insulating BSO thin films were deposited at $10^{-1}$ Torr and 800 °C onto (001)-oriented MgO single-crystal substrates and (001)-oriented 0.7PbMg$_{1/3}$Nb$_{2/3}$O$_3$–0.3PbTiO$_3$ (PMN-PT) single-crystal substrates. Subsequently, 12-nm-thick LaAlO$_3$ layers were grown at $10^{-6}$ Torr and 800 °C on top of the insulating BSO films.

The transmission electron microscopy characterizations were carried out by a FEI Talos F200X after the cross-section samples were obtained using focused ion beam technique. The ultraviolet-visible-infrared transmittance spectra of BSO//MgO heterostructures were collected by a SHIMADZU UV-3600 spectrometer at room temperature. The electrical measurements were conducted by a Quantum Design physical property measurement system with its resistivity option.

## 3. Results and discussion

To achieve that, we first fabricated 320-nm-thick BSO thin films onto (001)-oriented double-side-polished MgO single-crystal substrates at different oxygen partial pressures ranging from $10^{-1}$ to $10^{-7}$ Torr at 800 °C by pulsed laser deposition. Fig. 1(a) shows room-temperature sheet resistance of BSO thin films. The BSO films deposited at $10^{-3}$ Torr exhibit the lowest sheet resistance of 9.84 kΩ/□, while the films fabricated at a too high and a too low oxygen pressure are highly insulating. As pristine BSO is a large bandgap insulator, the appearance of the measurable conduction indicates the formation of point defects that have served as effective shallow donors for BSO. Generally, for oxide thin films, a high oxygen growth pressure leads to the formation of cationic vacancies while a low oxygen pressure results anionic vacancies. However, a too high or a too low oxygen pressure can also yield a large number of compensating defects, which could explain the oxygen-pressure-dependent conduction.

Fig. 1(b) plots the ultraviolet-visible-infrared transmittance spectra of BSO thin films collected at room temperature. The fitted refractive index via the thickness fringes is ~2.27, which is in good agreement with the refractive index 2.34 of bulk BSO. The fitted optical band using the indirect bandgap formula [21] is presented in Fig. 1(c), where BSO films deposited above $10^{-5}$ Torr exhibit a bandgap of ~3.17 eV and a high transmittance rate of more than 85% in the visible light range, comparable with bulk BSO. Nevertheless, BSO films fabricated at $10^{-6}$ and $10^{-7}$ Torr have both much lower bandgap and transmittance, suggesting the formation of another Ruddlesden-Popper phase of the barium stannate [22,23].

The measurable resistance for BSO films grown on MgO substrates at $10^{-3}$ and $10^{-4}$ Torr is less likely to be bandgap conduction as the bandgap is very large, above 3 eV. Instead, point defects such as cationic and anionic defects could serve as shallow electron donors, thus contributing to electrical conduction. Therefore, it is the number of defects that decide the measurable resistivity not the bandgap that does not change too much between $10^{-2}$ and $10^{-4}$ Torr.

We then focus on the conducting BSO//MgO heterostructures fabricated at $10^{-3}$ Torr. Due to the small lattice mismatch between MgO and BSO, the growth BSO on MgO is epitaxial as schematized in Fig. 1(d) and evidenced by transmission electron microscopy (TEM) characterization in Fig. 1(e & f). Temperature-dependent resistivity from 300 down to 50 K (Fig. 1(g)) reveals semiconductor-like behavior. Below 50 K, the resistance is too high and exceeds our measurement limit.

Hall measurements indicate the conductivity is *n*-type for all the temperatures ranging from 300 to 50 K, which suggests that the point defects in conducting BSO films are predominantly oxygen vacancies as they typical behave as electron donors. The carrier density is ~$5.84 \times 10^{19}$ cm$^{-3}$ at 300 K, characteristic of a semiconductor material, and slightly decreases to ~$5.08 \times 10^{19}$ cm$^{-3}$ at 50 K (Fig. 1(h)). The resulting electron mobility is rather low, less than 1 cm$^2 \cdot$V$^{-1} \cdot$s$^{-1}$, thus demonstrating the strongly localized nature of oxygen-vacancy-doped electrons in BSO films (Fig. 1(i)). The low mobility values and the temperature dependence of the mobility are consistent with the strong localization mechanism that becomes stronger at low temperatures.

To realize a well-confined 2DEG, BSO films deposited at $10^{-1}$ Torr that are highly insulating were utilized. Subsequently, a 12-nm-thick LAO layer was deposited on top of the BSO film. A cross-section TEM image of a LAO/BSO//MgO heterostructure is shown in Fig. 2(a), where zoom-in interfacial images are organized as insets. As revealed by TEM measurements, the BSO film is epitaxially grown on MgO and the BSO//MgO interface is sharp. In addition, the LAO/BSO interface as well as the crystalline nature of the LAO layer is clearly visible.

It was found that the deposition of a 12-nm-thick LAO top layer reduces the sheet resistance $R_S$ of a BSO film by more than 8 orders of magnitude to ~ 450 Ω/□ at room temperature (Fig. 2(b)). Furthermore, such a heterostructure presents metallic transport behavior, *i.e.*, $R_S$ decreases with reducing temperature, which reveals the formation of metallic conduction at the interface between LAO and BSO. Intriguingly, a resistance minimum emerges at ~80 K and $R_S$ increased by ~3% down to 10 K, as plotted in Fig. 2(b).

Hall measurements indicate that the conducting carriers are electrons at all the temperatures ranging from 300 to 10 K as well. Both the electrical conduction and the carrier density fully vanish after annealing in air at 600 °C for 1 h, which is consistent with the oxygen vacancy doping scenario [5,16]. The room-temperature sheet carrier density $n_S$ is ~7.72×$10^{14}$ cm$^{-2}$ (Fig. 2(c)), which is much higher than that of the oxygen-vacancy-induced 2DEG at the LAO/STO interface, ~1×$10^{14}$ cm$^{-2}$ [5] and polar-discontinuity-induced 2DEG at the LAO/STO interface, ~1.5×$10^{13}$ cm$^{-2}$ [5,24,25].

On the other hand, the possible polar discontinuity at the interface between non-polar BSO and polar LAO could contribute to the free carriers via the interfacial charge transfer [15]. In principle, this mechanism can yield a sheet carrier density of ~2.95×$10^{14}$ cm$^{-2}$ considering the BSO lattice constant 4.116 Å and the half electron transfer per uc. However, it is still much less than our experimental value. Moreover, the polar-discontinuity-induced 2DEG is robust over post oxygen annealing [5,24,25]. The intermixing at oxide interfaces fabricated at high temperature is inevitable [26,27], which can lead to La doping in BSO. Nevertheless, in this case, post annealing could further enhance the interdiffusion and hence increase the interfacial conductivity. Both cases are against our observation on annealing. Therefore, all the above results indicate the successful creation of an oxygen-vacancy-induced 2DEG at the LAO/BSO interface.

Furthermore, $n_S$ exhibits a weak temperature dependence and slightly decreases to ~$7.21\times10^{14}$ cm$^{-2}$ at 10 K (Fig. 2(c)). The fitted activation energy $\epsilon_a$ for carrier density via $n_S \propto e^{(-\epsilon_a/k_B T)}$ is ~0.97 meV, and is smaller than that of the oxygen-vacancy-induced 2DEG at the LAO/STO interface, ~4.2 meV [4]. This accounts for the absence of the carrier freeze-out effect, which has been extensively observed in oxygen-vacancy-doped STO [4,5,28,29]. Despite of the high carrier density, the mobility of the newly created 2DEG is remarkably high (Fig. 2(d)), ~18 cm$^2$·V$^{-1}$·s$^{-1}$ at 300 K as a result of the highly mobile Sn 5$s$ channel, which is one order of magnitude larger than that of STO-based 2DEG systems [2,4,5]. Although the room-temperature mobility value is lower than the highest values achieved in La-doped BSO thin films fabricated by molecular beam epitaxy [10-12], it could be further improved by enhancing the growth temperature [30] of BSO to increase the crystallinity and reduce disorder/defects.

High-field Hall measurements up to 14 T reveal a strictly linear Hall effect (Fig. 2(e)). This suggests that the LAO/BSO interface is dominated by single-band transport, which contrasts with the multiband transport in STO-based Ti 3$d$ interface systems [31-33], but consistent with the nature of the Sn 5$s$ orbit. Such a feature could be beneficial for enabling a high carrier mobility as it prohibits the interband scattering. Out-of-plane negative magnetoresistance is found for all the temperatures ranging from 5 to 150 K (Fig. 2(f)). It reaches ~-2.78% and ~-0.17% under 9 T at 5 and 150 K, respectively. Such a small, few percent of negative magnetoresistance is consistent with the weak localization mechanism [34]. As plotted in Fig. 2(g), all the magnetoconductance $\delta G = G(B) - G(0)$ data can be well fitted by the Hikami-Larkin-Nagaoka formula $\delta G = \frac{\alpha e^2}{2\pi^2 \hbar}\left[\psi\left(\frac{1}{2} + \frac{\hbar}{4eBl_\phi^2}\right) - \ln\left(\frac{\hbar}{4eBl_\phi^2}\right)\right]$ [35,36], where $\psi(x)$ is the digamma function, $l_\phi$ represents the phase coherence length and $\alpha$ is a material and temperature dependent parameter indicating the type of quantum conduction correction (weak localization $\alpha > 0$; weak antilocalization $\alpha < 0$).

The fitted parameter $\alpha$ is positive for all the temperatures and thus validates the weak localization mechanism, which is consistent with the observed low-temperature resistance minimum (Fig. 2(b)) as the constructive interference between time-reversed electron paths in weak localization increases the resistance. Meanwhile, the appearance of weak localization further supports the two-dimensional (2D) transport at the LAO/BSO interface [34]. $l_\phi$ is remarkably large at 5 K, ~48 nm, and rapidly decays with increasing temperature (Fig. 2(h)). Compared with the mean free path estimated from $l = \frac{h}{\sqrt{2\pi n_S e^2 R_S}}$ that is almost constant ~9 nm for all the temperatures, the critical temperature separating the dirty ($l < l_\phi$) and clean limit ($l > l_\phi$) for quantum conductance correction is found to be ~100 K.

To further investigate the nature of the conduction, we have carried out conducting atomic force microscopy measurements for a polished cross-section LAO/BSO//MgO sample. The cross-section surface was prepared by the standard method for transmission electron microscopy sample preparation, *i.e.*, the sample was cut into two pieces and pasted with glue face to face, and then mechanical polishing with diamond-coated discs and colloidal silica was performed to achieve a fine surface. As shown below, the most conducting channel (deep dark) residing in the BSO side is ~5 nm. It provides direct evidence for 2D transport (Fig. 3).

Furthermore, it was found that the room-temperature $R_S$ of a LAO/BSO//PMN-PT (0.7PbMg$_{1/3}$Nb$_{2/3}$O$_3$–0.3PbTiO$_3$) heterostructure is 340 times larger than that of a LAO/BSO//MgO heterostructure, reaching ~156 kΩ/□ (Fig. 4(a)). There is a lattice mismatch between bulk BSO ($a$ = 4.116 Å) and substrate materials including PMN-PT ($a$ = 4.02 Å) and MgO ($a$ = 4.203 Å). However, X-ray diffraction measurements found that epitaxial strain has largely relaxed (Fig. 4(b)) due to a large BSO film thickness of 320 nm. Despite of the same thickness, the (001) & (002) peaks of the BSO//PMN-PT thin film are much weaker than that of the BSO//MgO film, which indicates poor crystallinity of the BSO//PMN-PT film. That is mainly because Pb in PMN-PT is not stable and becomes volatile at high temperatures [37,38]. As a result, a large number of defects could be formed in BSO films during deposition, which lead to a much higher resistivity. Consequently, the temperature-dependent $R_S$ of the LAO/BSO//PMN-PT heterostructure exhibits semiconducting behavior.

Unlike the epitaxial strain that is static and irreversible and relaxes quickly from interfaces, piezoelectric strain triggered by electric fields in ferroelectric substrates such as PMN-PT is dynamic and reversible, and can be effective for thick films [39-41]. For example, via grazing-incidence X-ray diffraction measurements, Dekker *et al.* found that the in-plane piezoelectric strain transferred from PMN-PT is vertically homogeneous for a 600-nm-thick La$_{0.7}$Sr$_{0.3}$MnO$_3$/SrTiO$_3$ superlattice containing 100 oxide interfaces [39]. We therefore examined the piezoelectric strain effect via ferroelectric PMN-PT, which generates an in-plane biaxial compressive strain [37,41-51] due to non-180° ferroelastic polarization switching under a perpendicular electric field (Fig.

5(a)) [52]. As plotted in Fig. 5(b), electric-field-dependent $R_S$ presents two non-volatile states. At +4 kV/cm, the in-plane compressive strain generated from PMN-PT is ~0.08% [42], which results in a giant resistance enhancement of ~350% ($\Delta R_S$ ~ 545 kΩ/□).

Such a resistance modulation is less likely due to the electrostatic doping as a positive gating field should inject electrons into the 2DEG channel and hence lower the resistance. In addition, the static dielectric constant (~15) [53] of the BSO layer diminishes the electrostatic modulation of ferroelectric PMN-PT from ~30 μC/cm$^2$ to a negligible value of ~5.3 nC/cm$^2$ at 4 kV/cm. Indeed, the electric-field-dependent $R_S$ curve resembles the asymmetric strain curve originating from 109° ferroelectric domain switching in (001)-oriented PMN-PT [50,52,54] owing to the non-volatile strain states of PMN-PT. As a result, such a heterostructure can work as a memory device of ultralow power consumption (Fig. 5(d)) with a large resistance/voltage difference output. The strain modulation for the 2DEG at the surface of BSO is much larger than that for the three-dimensional resistance of a single-layer BSO film fabricated at 10$^{-3}$ Torr oxygen pressure (Fig. 6). This emphasizes the important role of quantum confinement in determining the strain sensitivity.

To understand the remarkable strain effect on the 2DEG, we have further carried out density functional theory calculations. It turns out that a compressive strain linearly increases the effective mass of the conduction band electrons in BSO while a tensile strain reduces the effective mass (Fig. 7). This is consistent with the recent theoretical calculations [55]. However, our theoretical modeling is based on a three-dimensional (3D) *bulk* BSO unit cell. For the 3D transport in BSO//PMN-PT heterostructures as shown in Fig. 6(c), the resistance modulation by piezoelectric strain (~0.1% compressive strain) triggered from PMN-PT is ~3%, which matches our theoretical modeling and calculation results (Fig. 7). In contrast, the 2DEG channel is much more sensitive to the strain (Fig. 5(b)). That is possibly because the band structure of a 2D electron system could be largely different from a 3D system due to multiple factors such as inversion symmetry breaking and quantum confinement. Therefore, the effective mass and its response to external stimuli such as strain could be significantly distinct. Our experimental results highlight the obvious advantage of the 2DEG channel for strain-related electronic device applications compared with the 3D conducting channel in multifunctional oxide BSO. More detailed band structure studies are needed for this 2DEG system.

## 4. Conclusions

In summary, we have developed an oxygen-vacancy-induced 2DEG based on the 5$s$ channel of BSO. Compared with STO-based Ti 3$d$ interface systems, the much less localized 5$s$ orbit and the absence of interband scattering ensure the high mobility at room temperature, and the much shallower donor level of oxygen vacancies leads to a greatly higher 2D carrier density and the absence of the carrier freeze-out effect. These features together with the quantum conductance correction and the ultrahigh strain sensitivity are rather useful for oxide electronic device applications at a broad temperature range. More importantly, such a 5$s$-orbit-based 2DEG provides a brand new and fertile ground for unearthing other exotic functionalities such as ferromagnetism and possible superconductivity and novel optical properties [4,56-65] at oxide interfaces.

**Acknowledgments**

Z.L. acknowledges financial support from the National Natural Science Foundation of China (NSFC; grant numbers 11704018, 51822101, 51861135104 & 51771009). Work at Fudan is supported by National Natural Science Foundation of China (11825403), the Special Funds for Major State Basic Research (2015CB921700), and Qing Nian Ba Jian Program. A portion of this work was performed on the Steady High Magnetic Field Facility, the High Magnetic Field Laboratory, Chinese Academy of Sciences. J.Y. acknowledges the support of Future Fellowship (Australia Research Council, FT160100205). D.Q. acknowledges the support of the Australian Research Council (Grant No. FT160100207). The authors would like to acknowledge the Singapore Synchrotron Light Source (SSLS) for providing the facility necessary for conducting the research. The Laboratory is a National Research Infrastructure under the National Research Foundation Singapore.

**Figure 1**

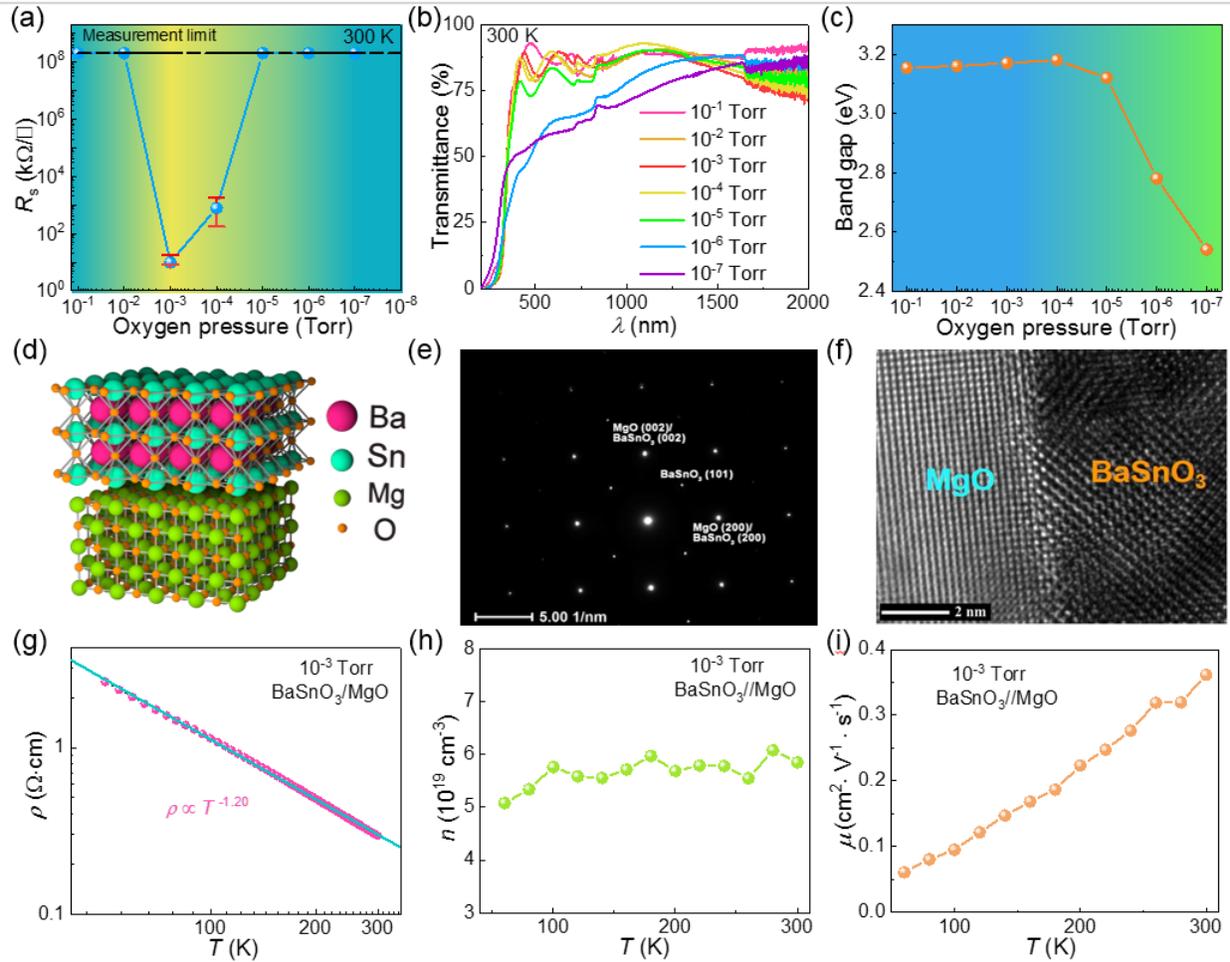

**Fig. 1.** Electrical, optical and structural properties of BaSnO$_3$ (BSO) thin films grown on (001)-oriented MgO substrates. (a) Room-temperature sheet resistance $R_S$ and (b) transmission spectra of BSO thin films grown at different oxygen partial pressure ranging from $10^{-1}$ Torr to $10^{-7}$ Torr. (c) Fitted band gaps of different BSO films according to the transmission spectra. (d) Schematic of epitaxial growth of BSO films onto MgO substrates. (e) Selected area diffraction pattern and (f) cross-section transmission electron microscopy image of a BSO//MgO heterostructure fabricated at $10^{-3}$ Torr. (g-i) Resistivity, carrier density and mobility versus temperature of the BSO//MgO heterostructure fabricated at $10^{-3}$ Torr oxygen partial pressure.

**Figure 2**

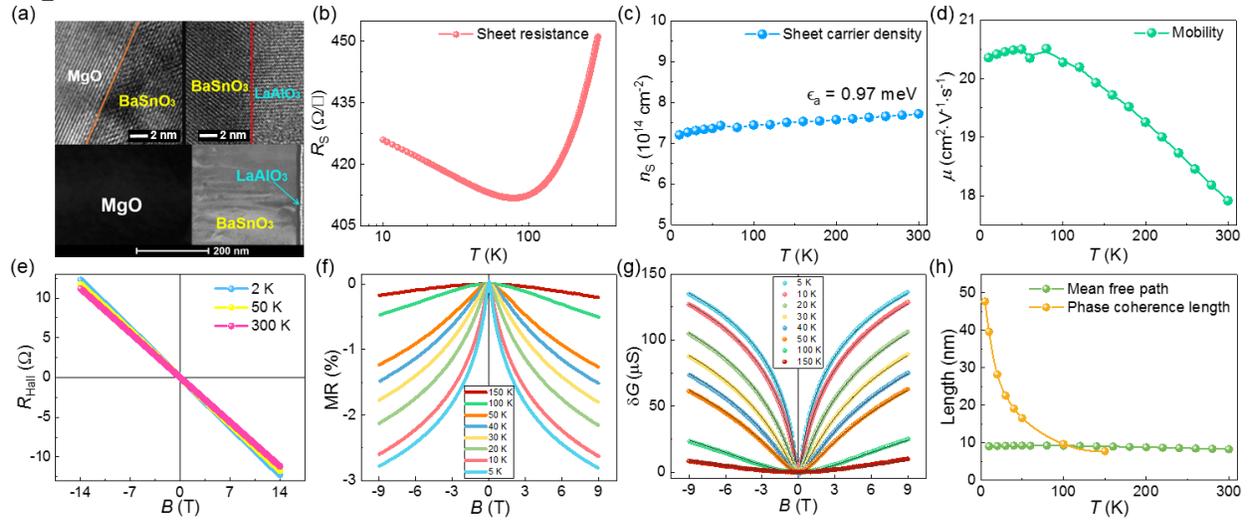

**Fig. 2.** Two-dimensional electron gas (2DEG) in a LaAlO$_3$/BSO//MgO (LAO/BSO//MgO) heterostructure. (a) Cross-section transmission electron microscopy images of a LAO/BSO//MgO heterostructure consisting of a 12-nm-thick LAO thin film and a 320-nm-thick BSO film. Left inset: An interfacial region of the BSO//MgO interface. Right inset: An interfacial region of the LAO/BSO interface. (b-d) Sheet resistance $R_S$, sheet carrier density $n_S$, and mobility $\mu$ as a function of temperature from 10 to 300 K, respectively. (e) Hall measurements at 2, 50 and 300 K up to 14 T. (f) Out-of-plane magnetoresistance up to 9 T at low temperatures ranging from 5 to 150 K. (g) Fitting of the magnetoconductance by the Hikami-Larkin-Nagaoka formula. Circle points represent experimentally measured magnetoconductance and grey solid lines are fitted curves. (h) Phase coherence length and mean free path versus temperature.

**Figure 3**

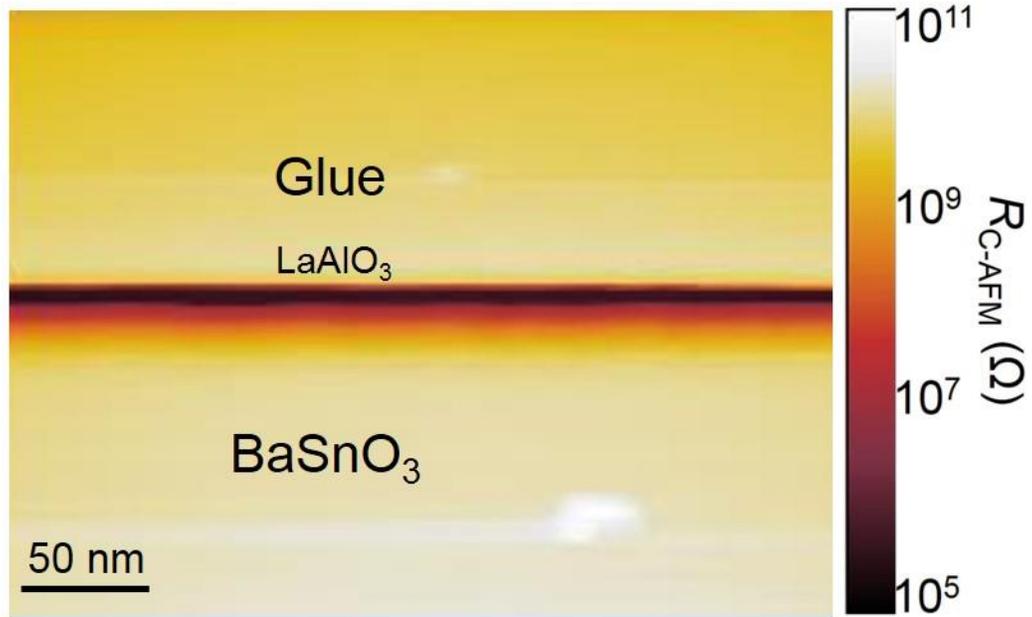

**Fig. 3.** Conduction mapping of the cross-section of a LAO/BSO//MgO heterostructure by conducting atomic force microscopy.

**Figure 4**

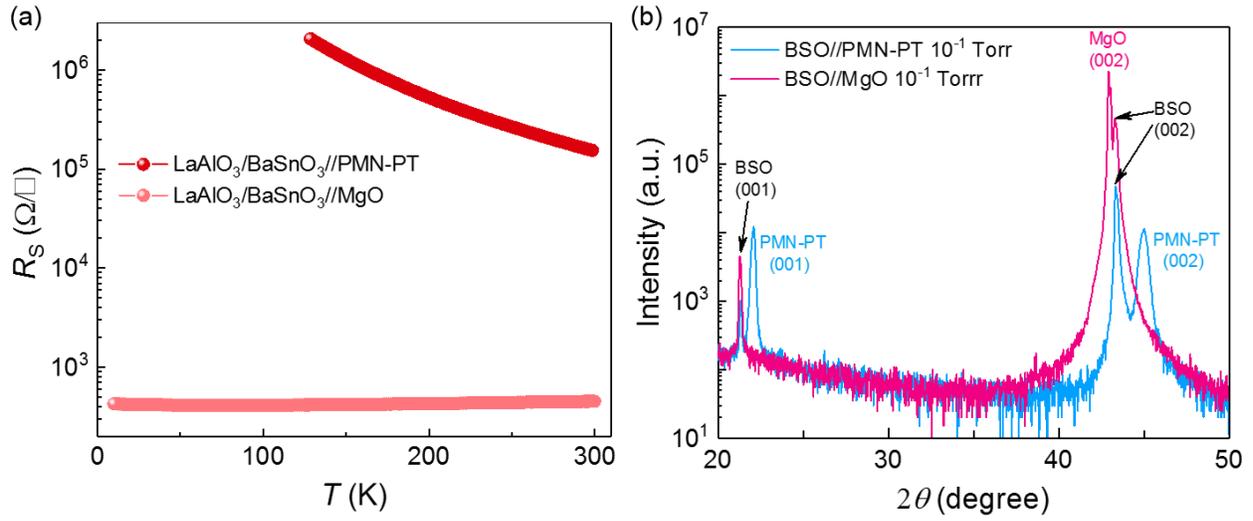

**Fig. 4.** (a) Temperature dependent sheet resistance $R_S$ of the LAO/BSO//MgO and the LAO/BSO//PMN-PT heterostructures. (b) X-ray diffraction patterns of the 320-nm-thick BSO films grown on MgO and PMN-PT at $10^{-1}$ Torr oxygen pressure.

Figure 5

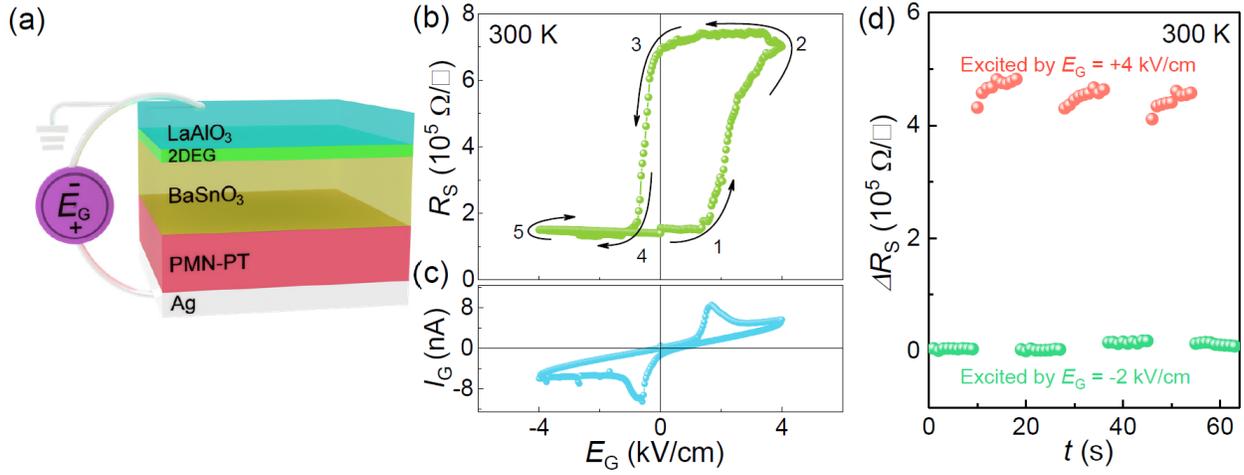

**Fig. 5.** Room-temperature piezoelectric strain modulation of the 2DEG at the LAO/BSO interface in a LAO/BSO//PMN-PT ($0.7PbMg_{1/3}Nb_{2/3}O_3$–$0.3PbTiO_3$) heterostructure. (a) Sketch of the measurement geometry of the electric-field gating via ferroelectric PMN-PT. (b) Electric-field-dependent sheet resistance of a LAO/BSO//PMN-PT heterostructure. The arrows and numbers indicate the measurement procedure. (c) Gating current of the PMN-PT versus the electric field. (d) Non-volatile high- and low-resistance states obtained by electric-field excitations of +4 kV/cm and -2 kV/cm, respectively.

**Figure 6**

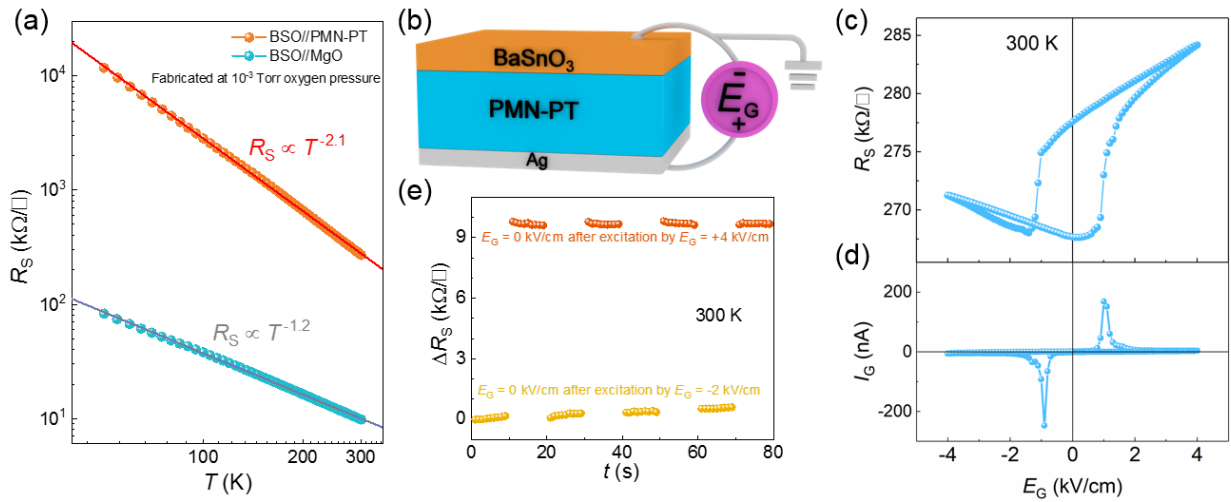

**Fig. 6.** Piezoelectric strain modulation of the BSO//PMN-PT heterostructures at room temperature. (a) Temperature dependent sheet resistance of the BSO//MgO and the BSO//PMN-PT heterostructures. (b) Schematic of the measurement geometry of the electric-field gating via ferroelectric PMN-PT. (c) Electric-field-dependent sheet resistance of a BSO//PMN-PT heterostructure. (d) Gating current of the PMN-PT versus the electric field. (e) Non-volatile high- and low-resistance states obtained by electric excitations of +4 kV/cm and -2 kV/cm, respectively.

**Figure 7**

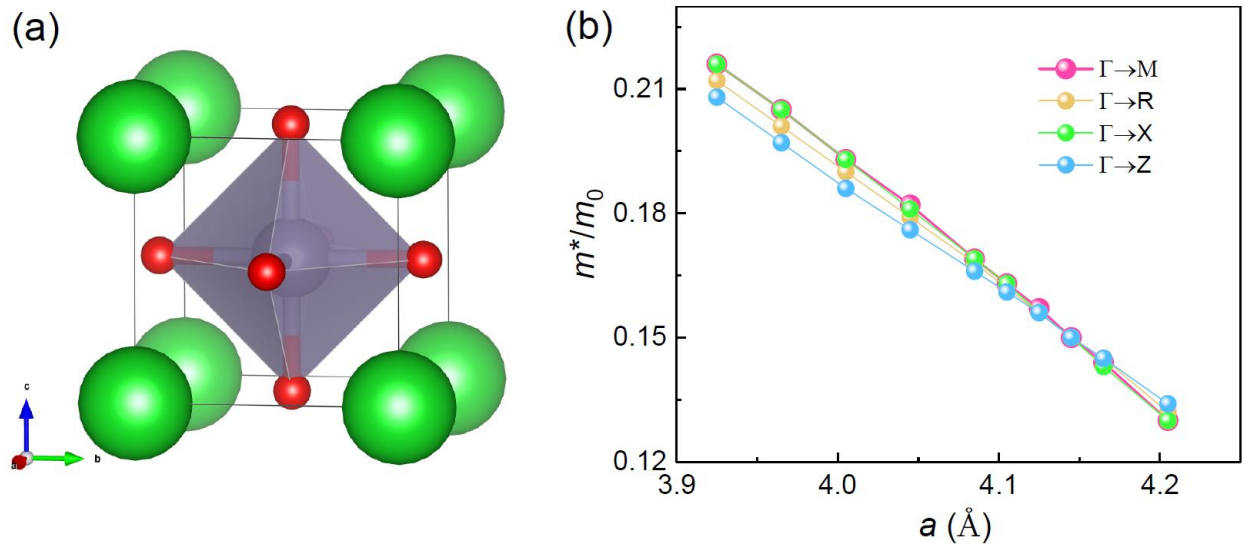

**Fig. 7.** Theoretical calculations. (a) Schematic of the crystal structure of a BSO unit cell. Green, red, and grey balls stand for Ba, O and Sn atoms, respectively. (b) Calculated effective mass of electrons in the conduction band of BSO as a function of its in-plane lattice constant along different directions in momentum space.